\documentclass[man]{apa7}
\usepackage[utf8]{inputenc}
\usepackage{amsmath}
\usepackage{graphicx}
\usepackage{listings}
\usepackage{booktabs}
\usepackage{makecell}
\usepackage{lscape}
\usepackage{rotating}

\usepackage[backend=biber, style=apa, natbib=true]{biblatex}
\usepackage{lineno}

\newcolumntype{P}[1]{>{\centering\arraybackslash}p{#1}}

\title{Designing PULSE: A Realtime Annotation Tool to Support Simulation Debriefing}
\shorttitle{Designing PULSE}

\authorsnames[1,2,3,4]{Caleb Vatral, Jennifer Hunt, Mary Ann Jesse, Gautam Biswas}
\authorsaffiliations{{Department of Computer Science, Tennessee State University, 3500 John A. Merritt Boulevard, Nashville, TN 37209, USA}, {School of Nursing, Vanderbilt University, 461 21st Ave S, Nashville, TN 37240, USA}, {School of Nursing, Colorado Mountain College, 802 Grand Avenue, Glenwood Springs, CO 81601, USA}, {Institute for Software Integrated Systems, Vanderbilt University, 1025 16th Ave S, Nashville, TN 37212, USA}}

\note{ ~\\
cvatral@tnstate.edu \\
jennifer.d.hunt@vanderbilt.edu \\
majessee@coloradomtn.edu \\
gautam.biswas@vanderbilt.edu
}

\authornote{
Corresponding Author: Caleb Vatral, cvatral@tnstate.edu \\
Funding: This research did not receive any specific grant from funding agencies in the public, commercial, or not-for-profit sectors.
}

\abstract{
\textbf{Background:} Debriefing is central to effective simulation-based education. However, effective debriefing is challenged by high instructor workloads and limited engagement of observing students. \\
\textbf{Methods:} A real-time annotation tool to support debriefing, called PULSE, was co-designed with nursing educators. A field study comparing three standard simulation debriefings with three debriefings using PULSE was conducted as a preliminary evaluation. Outcomes were assessed using the Debriefing Assessment for Simulation in Healthcare (DASH) student survey and a follow-up instructor interview. \\
\textbf{Results:} PULSE significantly improved overall DASH scores ($t(4)=4.03, p=0.027,$ Cohen's $d=2.05$). Survey findings suggested improvements in debriefing organization and depth of reflection. Interview data indicated that the student-generated annotations enhanced engagement and stimulated more interactive discussions. \\
\textbf{Conclusions:} PULSE shows promise as a support tool for debriefing, particularly by facilitating reflection of observing students. However, larger and more diverse studies are needed to confirm effectiveness and refine the system for broader implementation. \\ ~
}

\date{\today}

\begin{document}
\maketitle

\section{Introduction}
Simulation offers a safe, repeatable, and controlled environment for students to engage in experiential learning for healthcare education. Central to the effectiveness of simulation, the debriefing process allows students and instructors to engage in reflective dialogue after a simulation to solidify learning outcomes \citep{decker2025healthcare}. Effective debriefing promotes basic skill development, enhances critical thinking and reflection, and fosters the development of clinical reasoning. It is widely acknowledged as an essential component of effective and evidence-driven simulation-based education \citep{duff2024debriefing}.

Despite the recognized successes of simulation as a component of healthcare education and the importance of debriefing as part of simulation, challenges persist in ensuring best practice implementation. This study addresses two primary challenges in simulation-based education. 
\begin{enumerate}
    \item Instructors face a significant burden due to limited staffing and growing class sizes, which requires them to engage in multiple simulation roles simultaneously. These roles include acting as the simulated patient, managing the technical aspects of high-fidelity simulators, observing and assessing participating students to prepare for debriefing, and engaging students who are not participating in the simulation directly \citep{daug2019difficulties,park2025facilitators}.
    \item Despite evidence underscoring the valuable role that non-participating students can play in the collective learning experience, these observers are often not fully integrated into the simulation process \citep{turnbull2025learning,tutticci2022exploring}.
\end{enumerate}

In light of these challenges and following the recent successes of integrating technology tools into the debriefing process \citep{keiser2024systematic}, this paper presents a newly developed web application called \textit{PULSE} (\textbf{P}romoting \textbf{U}nderstanding and \textbf{L}earning through \textbf{S}imulation \textbf{E}vents), which was collaboratively designed with nursing educators. PULSE supports the debriefing process by enabling instructors and students to record their thoughts in real-time while a simulation evolves. Then, during debriefing, the PULSE dashboard allows for quick and easy navigation through these recorded events synced with video playback to support discussion and reflection.

To evaluate the new PULSE system, we conducted a field study with an experienced nursing instructor and her group of six nursing students across a set of standardized simulations. We compared baseline unmodified debriefing to debriefing utilizing PULSE using a within-subjects design, collecting both quantitative data from pre-validated student surveys and qualitative data from follow-up interviews. This data revealed a preference for debriefings utilizing PULSE and evidence suggesting that tools enabled enhanced structure of the debriefing, deeper reflective discussions, and increased student engagement.

Overall, this paper makes the following research contributions:
\begin{enumerate}
    \item The co-design and implementation of an interactive web application called \textit{PULSE} to discussion in simulation debriefing;
    \item Empirical data (both quantitative and qualitative) that provides evidence for the strengths and weaknesses of the new tools, which was generated from a field study comparing debriefing using PULSE to baseline debriefing methods; and
    \item New empirical data supporting the benefits of student-generated and student-guided discussion during debriefing.
\end{enumerate}

\section{Background}

\subsection{Simulation-Based Healthcare Education}
Within simulation-based education, debriefing is widely recognized as an essential component of the simulation process \citep{duff2024debriefing}. Debriefing is a structured discussion that takes place after a simulation event that allows students to reflect on their actions, decisions, and experiences during the simulation. During debriefing, a facilitator guides students through a discussion about what happened during the simulation, exploring students' feelings, knowledge, and decision-making \citep{decker2025healthcare, duff2024debriefing}. Studies across the literature consistently underscore the importance of debriefing in simulation-based education, with simulations that include debriefing demonstrating enhanced learning outcomes, improved clinical performance, and higher learner self-efficacy.\citep{lee2020debriefing,toqan2023effect,cao2026research}.

Various frameworks and tools have been developed to facilitate debriefing. These frameworks provide instructors with structured approaches to guide debriefing sessions, ensuring that key learning objectives are addressed and participants are actively engaged in the reflection process \citep{decker2025healthcare}. Examples of such frameworks include the PEARLS, Plus-Delta, and Debriefing for Meaningful Learning models, each offering similar structures combined with their own unique perspectives on debriefing methodology \citep{cheng2016promoting,cheng2021embracing,dreifuerst2012using}. In addition, some recent approaches have been incorporating technological aids --- such as video playback, performance recording systems, and learning analytics --- into the debriefing process, often successfully improving the experience and learning outcomes \citep{keiser2024systematic}.

\subsection{Current Issues in Simulation and Debriefing}
Despite the efficacy of simulation-based education, significant challenges remain in implementing simulation according to evidence-based best practices. Resources constraints, including a lack of staffing resources dedicated to the simulation process, often puts significant strain simulation infrastructure and faculty \citep{park2025facilitators,daug2019difficulties}.  
This is especially evident for instructors, who are often forced to play multiple roles during simulation sessions, including operating the manikin technology, acting as the simulated patient, and concurrently observing and assessing student performance in preparation for debriefing. Instructors may struggle to balance multiple responsibilities while maintaining engagement and providing meaningful feedback to students. In addition, these workload constraints also affect the ability of instructors to monitor and engage \textit{observing students} (i.e., those not directly involved in the simulation enactment). Recent literature highlights the benefits that engaging these observing students has on the collective educational experience \citep{o2020different,tutticci2022exploring,turnbull2025learning,husebo2024post}.

Motivated by these issues, this work presents an initial evaluation of a new technological aid for simulation and debriefing, which has been developed in close collaboration with educational stakeholders. By deploying tools that are specifically designed to address existing challenges and align with the preferences of instructors and administrators, we can optimize the utility of new tools while simultaneously reducing barriers and risks related to their adoption.

\section{Designing PULSE}

This study implemented a co-design process \citep{Sanders2008} to develop a new technological aid, called \textit{PULSE}. We began the design process by conducting semi-structured interviews aimed at understand the current practices and challenges in simulation. These interviews engaged two experienced faculty members who work directly with students and one assistant dean of the nursing school. Analysis of the interviews focused on identifying commonalities between participants' responses, which revealed a primary theme: multitasking challenges.

The instructors highlighted the significant challenges of simultaneously operating the simulated patient manikin, acting as (e.g., voicing) the patient, observing student performance, and preparing for debriefing. These multitasking challenges align with challenges highlighted in broader nursing education literature \citep{daug2019difficulties,park2025facilitators}. The instructors noted the difficulty of remembering all their observations that they wished to discuss during debriefing, often unable to take notes while also running the simulation. On instructor made an analogy to ambulatory monitoring, where a button can be pressed to mark an abnormality. She stated, ``\textit{I wish I had a button I could press during simulation to [quickly] mark an event like that}." 

This analogy was used as the basic design inspiration for the new tool. We engaged the instructors in a series of co-design activities to refine the concept and model the new tool's features and interface. We first performed a card-sorting activity \citep{alvarez2020deck}, which resulted in a model of the essential functionality of the application, centering on real-time marking of \textit{events} which can be recalled after the simulation on a dashboard to use as a discussion guide for debriefing. We then performed a \textit{dot-voting} activity \citep{verbert2020learning}, which allowed instructors to vote on specific features to prioritize in the prototype. Based on this, the researchers then created several low-fidelity paper prototypes \citep{sefelin2003paper} of the application, which were discussed with instructors. Their feedback was iteratively incorporated until a consensus was reached on a design for the new application. After consensus, a fully functional prototype of the new PULSE application was developed.

\section{The PULSE Application}\label{Paper1:sec:app}
PULSE is a progressive web application designed to aid the debriefing processes by allowing instructors and students to mark events of interest in real-time while watching a simulation. These marked events are displayed during debriefing on an interactive timeline-based video dashboard, enabling users to quickly review the events alongside a time-synced video recording of the simulation. The application has three basic components: (1) instructor view, (2) student view, and (3) debriefing dashboard.

\begin{figure}[t]
    \centering
    \includegraphics[width=\textwidth]{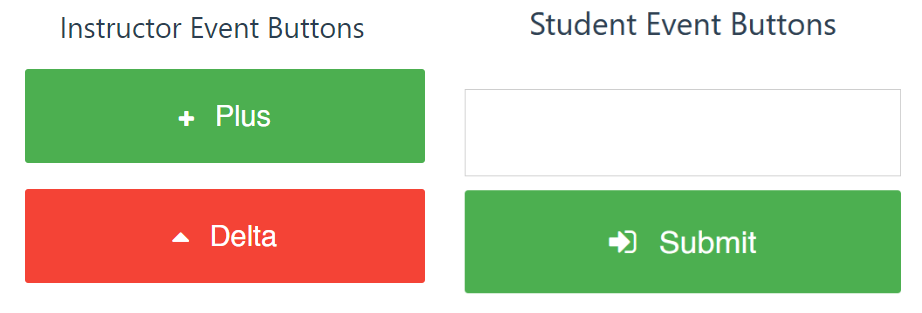}
    \caption{Event marking interfaces of the PULSE application for the instructor (left) and students (right)}
    \label{Paper1:fig:eventInterface}
\end{figure}

\subsection{Instructor Interface}
The instructor's event marking interface is depicted in Figure \ref{Paper1:fig:eventInterface} (Left). Upon initiating a new simulation session, the instructor works with a simplistic two-button interface for event marking. The buttons are color-coded to reflect the widely-used \textit{plus-delta} debriefing framework \citep{cheng2021embracing}, which categorizes simulation events into \textit{plus} (positive aspects) and \textit{delta} (areas for improvement). Instructors can mark plus and delta events to flag areas for discussion during debriefing. These marked events are later recalled on the debriefing dashboard. This process provides minimally invasive support for the instructor's memory, grounded in a flexible evidence-based debriefing methodology.

\subsection{Student Interface}
The student event interface is illustrated in Figure \ref{Paper1:fig:eventInterface} (Right). While an instructor runs the simulation and some students participate in the enactment, the remaining students observe the scenario via a live camera feed in the debriefing room. Similar to the instructor, these observers can mark events of interest in real-time while they watch. However, the student interface uses only a single, non-evaluative event button, avoiding categorizing events as positive or negative. Instead, the student-marked events are accompanied by text comments. The open-ended nature of these pairs of event and text is designed to encourage a wide range of discussion topics for debriefing, such as learner questions and thought processes, not just evaluation. In addition, to help mitigate students' potential hesitation to give negative feedback to their peers \citep{wong2022nursing} and potential fear of asking questions \citep{nadile2021call}, the student-marked events and comments are recorded and displayed anonymously. The use of these comments during debriefing enables greater engagement of observers and incorporation of more points of view in debriefing, which is consistent with recent evidence and established best practice standards \citep{decker2025healthcare, tutticci2022exploring}.

\subsection{Debriefing Dashboard}\label{Paper1:subsec:dashboard}

\begin{figure}[t]
    \centering
    \includegraphics[width=\textwidth]{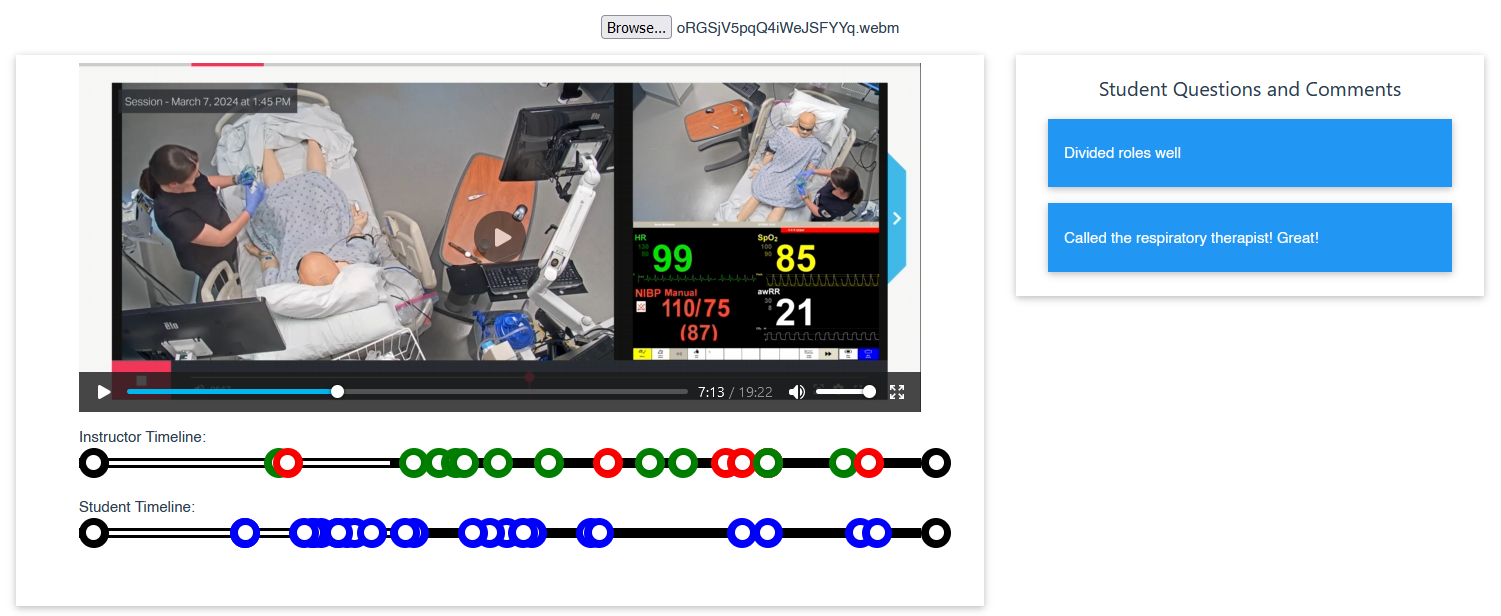}
    \caption{Debriefing dashboard of the PULSE application}
    \label{Paper1:fig:debrief}
\end{figure}

The debriefing dashboard is shown in Figure \ref{Paper1:fig:debrief}. This view is designed for the instructor to use after the simulation enactment to guide the debriefing. It contains three essential elements: 
\begin{enumerate}
    \item In the top left-hand area is the video player, which displays a recorded video of the simulation. Playback of these recorded videos enables accurate recall of the simulation and encourages debriefing discussions to be rooted in observable behaviors, consistent with established best practices \citep{decker2025healthcare}.
    \item Below this video are the timelines of events marked by the instructor and students. Each dot represents a single marked event, and the instructor can click on any event to advance the video player to the associated timestamp. This allows the instructor to quickly recall their own and their students' observations to support discussion and relevant video playback. This quick navigation also supports the integration of video in debriefing when time is limited, a problem identified in the INASCL Standards of Best Practice \citep{decker2025healthcare}.
    \item On the right-hand side of the video and timeline elements, the comments associated with each student-marked event are displayed. As the video plays or the instructor skips around in the video, this comment element dynamically updates to show only the comments marked within a small time range of the current timestamp, thereby allowing the group to easily see the comments which are relevant to a specific section of the simulation.
\end{enumerate}

\section{Field Study Methods}\label{Paper1:sec:methods}
To evaluate the usability of the PULSE system and its efficacy for enhancing debriefing, we conducted a mixed-methods within-group field study comparing baseline unmodified debriefing to debriefing utilizing PULSE. One experienced nursing faculty member and her university-assigned class group of six pre-licensure nursing students from a convenience sample in a university nursing program in the southeastern United States participated in the study, which followed them through six standardized adult-care clinical simulations using Laerdal high-fidelity manikins. First, they performed three baseline simulations with no changes to the standard simulation and debriefing procedures employed by the university. The instructor used a semi-structured Plus-Delta debriefing method for baseline. After the baseline simulations, researchers introduced the PULSE tools to the group, and they were used for the remaining three simulations. No changes to the simulation curriculum were made by the researchers other than the introduction of PULSE. Student participants ranged in age from 22 to 41 years old (M$=27$, Mdn$=24$, SD=$7$). Other specific demographic data was not collected from the participants, but the university nursing program from which the sample was taken is comprised of $90\%$ female students and $26\%$ minority students. The instructor was a faculty member at the university with over 10 years of teaching experience and held DNP, APRN, and FNP-C credentials. To reduce bias, she was not involved with the design activities of PULSE previously described. The study was approved by the Vanderbilt University institutional review board and all participants (students and instructors) provided written informed consent. 

The students filled out the \textit{Debriefing Assessment for Simulation in Healthcare} (DASH; student version) \citep{brett2012debriefing}, to evaluate their experiences in debriefing. DASH comprises six elements. Element one evaluates pre-briefing. The study design included only one pre-briefing before any of the simulations were conducted, so data for this element was not collected or evaluated. Elements two through six are listed in Table \ref{Paper1:tab:SurveyResults}. Each element is scored on a 7-point scale (1 = extremely ineffective to 7 = extremely effective). Students completed the survey twice: once after the three control simulations and once after the three experimental simulations. They were instructed to evaluate their experience for only the previous three simulations. A follow-up semi-structured interview was conducted with the instructor one week after the study to discuss her experiences using PULSE. Student participants were also invited for follow-up interviews, but none responded, likely due to the high demands of their coursework in the nursing program.

The DASH survey results were analyzed using paired t-tests to determine the significance of effects on debriefing by introducing PULSE. Our primary hypothesis was: 
\begin{quotation}
\textbf{H1:} \textit{Debriefing with PULSE will be more effective than the control condition, as measured by the overall DASH score, Student Version.} 
\end{quotation}
To evaluate this hypothesis, a paired t-test was used to compare baseline and experimental scores from the DASH survey. In addition, exploratory analyses using paired t-tests and effect size calculations for each DASH element were perfored to evaluate more specific element-level changes. These analyses were executed on the complete dataset ($n=6$) and a refined subset excluding outliers ($n=4$). A rating was considered an outlier if the student gave the maximum possible score for every element in both the control and experimental conditions. The instructor's follow-up interview was transcribed and analyzed by the first author, cross-referencing with student survey data for triangulation.

\section{Results}
\label{Paper1:sec:results}

\subsection{Statistical Analysis of DASH Surveys}
Table \ref{Paper1:tab:SurveyResults} shows the mean and standard deviation for each survey category for the control (ctrl) and experimental (exp) conditions, along with the results of the paired t-test and effect sizes for the full dataset (F) and the dataset with outliers removed (O). The results suggest that PULSE improved the overall quality of the debriefing experience as measured by DASH, with $t(4) = 4.030, p=0.027$ and a Cohen's $d$ effect size of $d=2.046$. 

Further exploratory analysis revealed smaller effect sizes for individual DASH elements compared to the overall score. This was expected given the low statistical power of the pilot study and the potential ceiling effects observed in DASH scores. However, certain elements indicated some enhancements to debriefing, albeit with weaker statistical significance. Survey data suggested an improved debriefing structure ($p=0.078, d=1.317$), deeper discussions ($p=0.068, d=1.398$), and better identification of strengths and areas for improvement ($p=0.066, d=1.414$). These modest individual improvements produced a larger effect when aggregated in the overall survey score.

\begin{sidewaystable}
    \renewcommand{\arraystretch}{0.5}
    \centering
    \begin{tabular}{p{0.35\textwidth}P{0.03\textwidth}P{0.05\textwidth}P{0.05\textwidth}P{0.05\textwidth}P{0.05\textwidth}P{0.06\textwidth}P{0.08\textwidth}P{0.1\textwidth}}
        \toprule
       \makecell{\textbf{~DASH Survey Item~}} & 
       \textbf{F/O} & 
       \multicolumn{2}{c}{\textbf{~Mean~}} & 
       \multicolumn{2}{c}{\textbf{~Std~}} &
       \textbf{~t-Stat~} & \textbf{~p-Value~} & \textbf{~Cohen's d~} \\
       \cmidrule(lr){3-4}
       \cmidrule(lr){5-6}
         & & Ctrl & Exp & Ctrl & Exp & & & \\
       \midrule[\heavyrulewidth]
       
       The instructor maintained an engaging context for learning. & \makecell{F \\ O} & 
       \makecell{6.700 \\ 6.550} & \makecell{6.800 \\ 6.700} & \makecell{0.329 \\ 0.300} & \makecell{0.310 \\ 0.346} & \makecell{1.000 \\ 1.000} & \makecell{0.363 \\ 0.391} & \makecell{0.408 \\ 0.500} \\ \midrule
       
       The instructor structured the debriefing in an organized way. & \makecell{F \\ O}  & 
       \makecell{6.333 \\ 6.000} & \makecell{6.708 \\ 6.563} & \makecell{0.626 \\ 0.456} & \makecell{0.332 \\ 0.315} & \makecell{2.087 \\ 2.635} & \makecell{0.091 \\ 0.078} & \makecell{0.852 \\ 1.317} \\ \midrule
       
       The instructor provoked in-depth discussions that led me to reflect on my performance. & \makecell{F \\ O}  & 
       \makecell{6.189 \\ 5.783} & \makecell{6.878 \\ 6.817} & \makecell{0.827 \\ 0.694} & \makecell{0.142 \\ 0.137} & \makecell{2.161 \\ 2.795} & \makecell{0.083 \\ 0.068} & \makecell{0.880 \\ 1.398} \\ \midrule
       
       The instructor identified what I did well or poorly — and why. & \makecell{F \\ O}  & 
       \makecell{6.250 \\ 5.875} & \makecell{6.917 \\ 6.875} & \makecell{0.822 \\ 0.750} & \makecell{0.204 \\ 0.250} & \makecell{2.169 \\ 2.828} & \makecell{0.082 \\ 0.066} & \makecell{0.886 \\ 1.414} \\ \midrule
       
       The instructor helped me see how to improve or how to sustain good performance & \makecell{F \\ O}  & 
       \makecell{6.500 \\ 6.250} & \makecell{7.000 \\ 7.000} & \makecell{0.658 \\ 0.687} & \makecell{0.000 \\ 0.000} & \makecell{1.859 \\ 2.183} & \makecell{0.122 \\ 0.117} & \makecell{0.760 \\ 1.091} \\ \midrule
       
       \makecell{\textbf{Overall}} & \makecell{F \\ O}  & 
       \makecell{6.401 \\ 6.108} & \makecell{6.843 \\ 6.776} & \makecell{0.560 \\ 0.410} & \makecell{0.199 \\ 0.192} & \makecell{2.538 \\ 4.030} & \makecell{0.052 \\ 0.027} & \makecell{1.036 \\ 2.046} \\
       
       \bottomrule
    \end{tabular}
    \caption{Statistical analysis results of the DASH student survey data collected at the end of each condition. \textit{F} indicates the full dataset, \textit{O} indicates the dataset with outliers removes, \textit{Ctrl} indicates the control condition, and \textit{Exp} indicates the experimental condition.}
    \label{Paper1:tab:SurveyResults}
\end{sidewaystable}

\subsection{Follow-Up Interview Analysis}
The instructor follow-up interview provided insights into the impact of PULSE on the debriefing process from the perspective of the nursing instructor. Much of the interview discussion focused on the impact of student-generated comments. The instructor expressed excitement over seeing observer-generated comments and their potential impact on the interactivity of the debriefing discussions. She stated,
\begin{quote}
    ``\textit{Yeah, I think capitalizing on what they observed. And what they noted makes it more interactive and makes it more where they are initiating their observation skills and noticing things that their peers did.}''
\end{quote}

The instructor went on to discuss how this greater interactivity increases the depth of debriefing discussions. She noted that including student perspectives and insights helped to foster more impactful exchanges, stating,
\begin{quote}
    ``\textit{Because it's experiential learning, all their learning is based on their experience or their interpretation of the experience. So when it was student-driven, I loved that part of it, because that's what solidifies things for them. Things that they wouldn't pick up if I just taught them, but things that they pick up on because they've experienced it themselves, or they've seen a peer experience it. So I think anything I try to teach after that, if they can base that on an experience, then it tends to stick with them longer.}''
\end{quote}

This feedback is consistent with the student DASH survey results, which indicated that debriefings that utilized PULSE ``\textit{provoked in-depth discussions that led me to reflect on my performance}" (DASH Element 4) and ``\textit{identified what I did well or poorly --- and why}'' (DASH Element 5) better than the baseline debriefings.

The instructor also shared some challenges with her adoption of PULSE. She reported that she wished she had marked fewer events than she did during the study, as it was difficult for her to keep track of everything she marked and prioritize when it came time for debriefing.
\begin{quote}
    ``\textit{I think I got clicker happy. I got, I got a little too into it. Like, oh, I need to say something about that. Oh, I need to say something about that. So I think by the time I got to the [debriefing] room, it was like, Oh, wow, I clicked on like 10 different things. I don't remember the first three.}''
\end{quote}
The instructor later indicated that additional training on the PULSE tools and more practice with simulations built around utilizing them would likely improve the process in the future.

\section{Discussion}\label{Paper1:sec:discussion}
The results of this field study underscored PULSE's impact on the structure, engagement, and reflective depth of simulation debriefings. Of particular note was the value of student-generated comments. The instructor seemed to value PULSE's student comment features more than the instructor-facing plus-delta interface. This was unexpected, given the emphasis placed on the instructor's event marks during the initial design process of PULSE. However, this result points toward the value of including more student-guided discussions in debriefings, particularly by engaging observing students. This idea is supported by a growing body of evidence \citep{tutticci2022exploring}, and it aligns with best practice standards for incorporating multiple points of view in debriefing \citep{decker2025healthcare}. 

One way to capture this observer perspective is to utilize peer debriefing methods, where observing students facilitate the debriefing instead of instructors. However, evidence surrounding the efficacy of peer-led debriefing is mixed \citep{khan2026effectiveness,arabi2023perceptions}, with some authors noting that learning outcomes could be impacted by inaccurate information due to a peer’s lack of expertise \citep{decker2025healthcare}. Hybrid facilitation methods are emerging as an alternative that still integrates observer feedback while limiting the drawbacks of peer-led debriefing. Many of these hybrid approaches use multiple phases of debriefing, with some phases led by students and others by instructors \citep{rueda2021combination,husebo2024post,fenzi2025expository}. 

Other hybrid techniques are still primarily run by instructors, but introduce structuring tools designed to engage observers during the simulation and guide them toward generating observations and feedback that can be discussed during debriefing \citep{turnbull2025learning,wighus2018educational}. This during-simulation elicitation approach bears a strong similarity to the design of PULSE, and studies that employ this technique show similar results to our results, surrounding ideas of enhanced student engagement and reflection. We hypothesize that PULSE's effectiveness stems primarily from its ability to increase the engagement of observers. This increased engagement moves debriefing discussions toward a hybrid approach that is still facilitated by an instructor but centered more on student-generated observations.

\subsection{Limitations and Future Research}\label{sec:limitations}
This work had several limitations that should be addressed in future studies. As an initial pilot study for PULSE, our sample size was small and lacked diversity, so researchers should be careful not to over-generalize our results. Given the successes of this small pilot study, follow-up studies will incorporate more participants from a wider subject pool. Since students rated baseline debriefings consistently high on the DASH scales, a ceiling effects in the DASH survey data may have further impacted the statistical analysis. Future work should consider using an independent trained rater to evaluate debriefings (e.g., DASH Rater Version). Finally, using a within-subjects study design may have influenced results due to ordering and learning effects, differences between simulation content in the control and experimental conditions, or participant response biases. A between-subjects design, where control and experimental conditions use the same instructors and simulations, may help to mitigate these potential issues.

Our results present significant opportunities for continuing research. We plan to explore how to maximize the potential of student-generated comments as a key feature of PULSE, as well as how PULSE can be adapted to support various structured debriefing methodologies other than Plus-Delta (e.g., PEARLS, DML, etc.). In addition, there are no elements of PULSE’s design that are specific to nursing simulation, so further studies should explore how PULSE and similar tools could be applied to other fields that utilize simulation-based education. Beyond PULSE, the results of this work point toward the need for expanded research examining the impact of student-generated discussion topics and hybrid facilitation models on debriefing efficacy.

\section{Conclusion}
This paper presented the design and evaluation of PULSE — a web application that enhances simulation debriefing by providing tools for the collection and review of thought from instructors and observing students in real-time during a simulation. A mixed-methods field study comparing baseline debriefing methods to debriefing utilizing PULSE highlighted the impact of PULSE on the structure, engagement, and depth of debriefing discussions. The inclusion of observing student comments that can be used to guide debriefing discussions was shown to be of critical importance to these improvements. By improve students’ engagement and critical reflection skills in simulation, we can empower the next generation of professionals to meet modern healthcare’s complex challenges.


\printbibliography

\end{document}